\documentclass[a4paper,11pt]{article}
\usepackage{jinstpub} % for details on the use of the package, please see the JINST-author-manual
\usepackage{lineno}

\title{\boldmath Reconstruction of neutrino events in the Accelerator Neutrino Neutron Interaction Experiment: Part I}

\author[a]{S. Abubakar,}
\author[b]{M. Acsencio-Sosa,}
\author[c]{A. Augusthy,}
\author[d]{D. Ajana,}
\author[d]{M. A. Aman,} 
\author[e]{J. Beacom,}
\author[f]{M. Bergevin,}
\author[g]{D.Bick,}
\author[h]{M. Breisch,}
\author[i]{G. Caceres Vera,}
\author[f]{S. Dazeley,}
\author[b]{S. Doran,}
\author[j]{E. Drakopoulou,}
\author[b]{S. Edayath,}
\author[k]{R. Edwards ,}
\author[l]{J. Eisch,}
\author[m]{N. Everitt,}
\author[b]{Y. Feng,}
\author[l]{V. Fischer,}
\author[i]{D. Fleming,}
\author[n]{R. Foster,}
\author[l]{S. Gardiner,}
\author[i]{B. Gelli,}
\author[c]{N. Goehlke,}
\author[o]{A. Gupta,}
\author[i]{P. Hackspacher,}
\author[g]{C. Hagner,}
\author[i]{J. He,}
\author[g]{B. Kaiser,}
\author[a,u]{M. Kandemir,}
\author[i]{C. Karagiannis,}
\author[h]{T. Lachenmaier,}
\author[m]{F. Lemmons,}
\author[b]{F. Krennrich,}
\author[n]{M. Malek,}
\author[c]{J. Martyn,}
\author[p]{A. Mastbaum,}
\author[c]{D. Maksimovic,}
\author[l]{C. McGivern,}
\author[p]{J. Minock,}
\author[d]{L. Mora-Lepin,}
\author[p]{C. Nguyen,}
\author[c]{M. Nieslony,}
\author[k]{M. O'Flaherty,}
\author[q,r]{G. D. Orebi Gann,}
\author[a]{B.K. Ozdemir,}
\author[i]{E. Pantic,}
\author[f]{T. Pershing,}
\author[q,r]{L. Pickard,}
\author[o]{N. Poonthottathil,}
\author[b]{E. Pottebaum,}
\author[k]{B. Richards,}
\author[s]{R. Rosero,}
\author[b]{H. Sogarwal,}
\author[d]{M. Sanchez,}
\author[c]{D. Schmid,}
\author[t]{M. Smy,}
\author[g]{M. Stender,}
\author[d]{A. Sutton,}
\author[i,1]{R. Svoboda,\note{Corresponding author.}}
\author[b]{C. Sweeney,}
\author[a,b]{E. Tiras,}
\author[t]{M. Vagins,}
\author[b]{V. Veeraraghavan,}
\author[m]{J. Wang,}
\author[b]{M. Wetstein,}
\author[b]{A. Weinstein,}
\author[c]{M. Wurm,}
\author[s]{M. Yeh,}
\author[i]{T. Zhang}

\affiliation[a]{Erciyes University, Department of Physics, Kayseri, 38030, T\"urkiye}
\affiliation[b]{Iowa State University, Department of Physics and Astronomy, Ames, IA 50011 U.S.A.}
\affiliation[c]{Johannes Gutenburg Universit\"at, Institut f\"ur Physik, Mainz 55128, Germany}
\affiliation[d]{Florida State University, Department of Physics, Tallahassee, FL 32306 U.S.A.}
\affiliation[e]{The Ohio State University, Department of Physics, Columbus, OH 43210 U.S.A.}
\affiliation[f]{Lawrence Livermore National Laboratory, Livermore, CA 94550 U.S.A.}
\affiliation[g]{Universit\"at Hamburg, Institut f\"ur Experimentalphysik, Hamburg 22761, Germany}
\affiliation[h]{Eberhard Karls Universit\"at, Kepler Center for Astro and Particle Physics, T\"ubingen 72076, Germany}
\affiliation[i]{University of California at Davis, Department of Physics and Astronomy, Davis, CA 95616, U.S.A.}
\affiliation[j]{N.C.S.R. “Demokritos,” Institute of Nuclear and Particle Physics, Agia Paraskevi 15341, Greece}
\affiliation[k]{University of Warwick, Department of Physics, Coventry CV4 7AL U.K.}
\affiliation[l]{Fermi National Accelerator Laboratory, Batavia, IL 60510, U.S.A.}
\affiliation[m]{South Dakota School of Mines and Technology, Physics Department,  Rapid City SD, 57701 U.S.A.}
\affiliation[n]{University of Sheffield, Department of Physics and Astronomy, Sheffield, S10 2TN, U.K.}

\affiliation[o]{Indian Institute of Technology , Department of Physics, Kanpur 208016, India}
\affiliation[p]{Rutgers University, Piscataway, NJ 08854 U.S.A.}
\affiliation[q]{University of California, Berkeley, Physics Department, Berkeley, CA 94720 U.S.A.}
\affiliation[r]{Lawrence Berkeley National Laboratory, Nuclear Science Division, Berkeley, CA 94720 U.S.A.}
\affiliation[s]{Brookhaven National Laboratory, Upton, NY 11973, U.S.A}
\affiliation[t]{University of California at Irvine, Department of Physics and Astronomy, Irvine CA, 92697 U.S.A.}
\affiliation[u]{Recep Tayyip Erdogan University, Department of Physics, Rize, 53100, T\"urkiye}

\emailAdd{rsvoboda@physics.ucdavis.edu}

\abstract{The Accelerator Neutrino Neutron Interaction Experiment (ANNIE) was designed to reconstruct neutrino events from the Fermilab Booster Neutrino Beam (BNB) with the parallel goals of measuring neutron production in interactions with oxygen and serving as a testbed for new technology. The ANNIE detector consists of a 26-ton water Cherenkov target tank instrumented with conventional photomultiplier tubes (PMTs), a downstream tracking muon spectrometer , and an upstream double wall of plastic scintillator to serve to veto charged particles incoming from neutrino events that occur upstream of the experimental setup. ANNIE has also deployed multiple Large-Area Picosecond PhotoDetectors (LAPPDs) and a test vessel of water-based liquid scintillator (WbLS). This paper describes the event reconstruction performance of the detector before implementation of these novel technologies, which will serve as a baseline against which their impact can be measured.  That said, even the techniques used for event reconstruction using only the conventional PMT array and muon spectrometer are significantly different than those used in other water Cherenkov detectors due to the small size of ANNIE (which makes nanosecond-scale timing not as useful as in a large detector) and the availability of reconstruction information from the tracking muon spectrometer. We demonstrate that combining the information from these two elements into a single fit using only pattern recognition yields a muon vertex uncertainty of 60 cm, a directional uncertainty of 13.2 degrees,  and energy reconstruction uncertainty of about 10\% for BNB muon neutrino Charged Current Zero Pion (CC0$\pi$) events.}


\begin{document}
\maketitle
\flushbottom

\section{Introduction}\label{sec:intro}

Water Cherenkov detectors have been in use for many years as a practical way to  track secondary charged particles from neutrino interactions in a large target mass for a relatively low cost. Underground detectors up to 50 kilotons have been built~\cite{IMB, SuperK} and the Hyper-Kamiokande detector~\cite{HyperK} will have a mass of over 250 kilotons. An advantage of such detectors is that they are sensitive to neutron capture gammas~\cite{SKneutron, SNOPneutron}, especially if the water is loaded with a capture agent such as gadolinium~\cite{egads, annie, skgd}. Understanding the neutrino-induced neutron multiplicity in interactions is important in that such neutrons can carry away significant energy from the primary vertex, and in addition they carry information on the initial neutrino interaction itself. 
In this paper we describe the vertex reconstruction and tracking performance of the Accelerator Neutrino Neutron Interaction Experiment (ANNIE), which was built to measure the neutron multiplicity from muon neutrino interactions on oxygen in the Fermilab Booster Neutrino Beam (BNB) and also to demonstrate the impact on neutrino event reconstruction of two new technologies: (i) Large Area Picosecond PhotoDetectors (LAPPDs) and (ii) Water-based Liquid Scintillator (WbLS).

The following sections describe the basic elements of the ANNIE detector, which includes (i) a target tank containing 26 tons of gadolinium-loaded water and a conventional PhotoMultiplier Tube (PMT) array, (ii) a plastic scintillator and iron plate muon spectrometer known as the Muon Range Detector (MRD), and (iii) an upstream plastic scintillator array called the Front Muon Veto (FMV). The performance described here is based only the conventional PMT array, the FMV, and the MRD as applied to muon track and energy reconstruction for Charged Current Zero Pion (CC0$\pi$) events from BNB muon neutrino interactions occurring in the target tank. In later papers we will describe the improvements in this baseline performance enabled by use of LAPPDs~\cite{annie_lappd} which have very fast timing (60 -- 70 ps) compared to the PMTs deployed here, and of WbLS~\cite{annie_wbls} which allows for the detection of charged particles below Cherenkov threshold in addition to improving the light yield (and hence detectability) of neutron capture events following neutrino interactions.

Note that even using these conventional beam elements the small size of ANNIE does not allow for the mostly time-based reconstruction used in much larger Cherenkov detectors, as the PMT timing uncertainty of 1--2 ns ($\sigma$) due to Transit Time Spread (TTS) is large compared to the PMT array size (2.3 m diameter). This paper describes novel techniques that were developed to combine tank PMT Cherenkov pattern recognition with MRD data to reconstruct the track and energy of neutrino-induced muons with minimal use of PMT timing.
 The neutron detection capabilities of ANNIE are also not utilized here and thus will be the subject of a future paper. The goal of this paper is to describe the first steps towards the ultimate goal of ANNIE, which is to develop the technology and reconstruction techniques for future large optical detectors.

\section{Overview of the ANNIE Detector}

\begin{figure}[h!]
	\begin{center}
\includegraphics[width=0.95\linewidth]{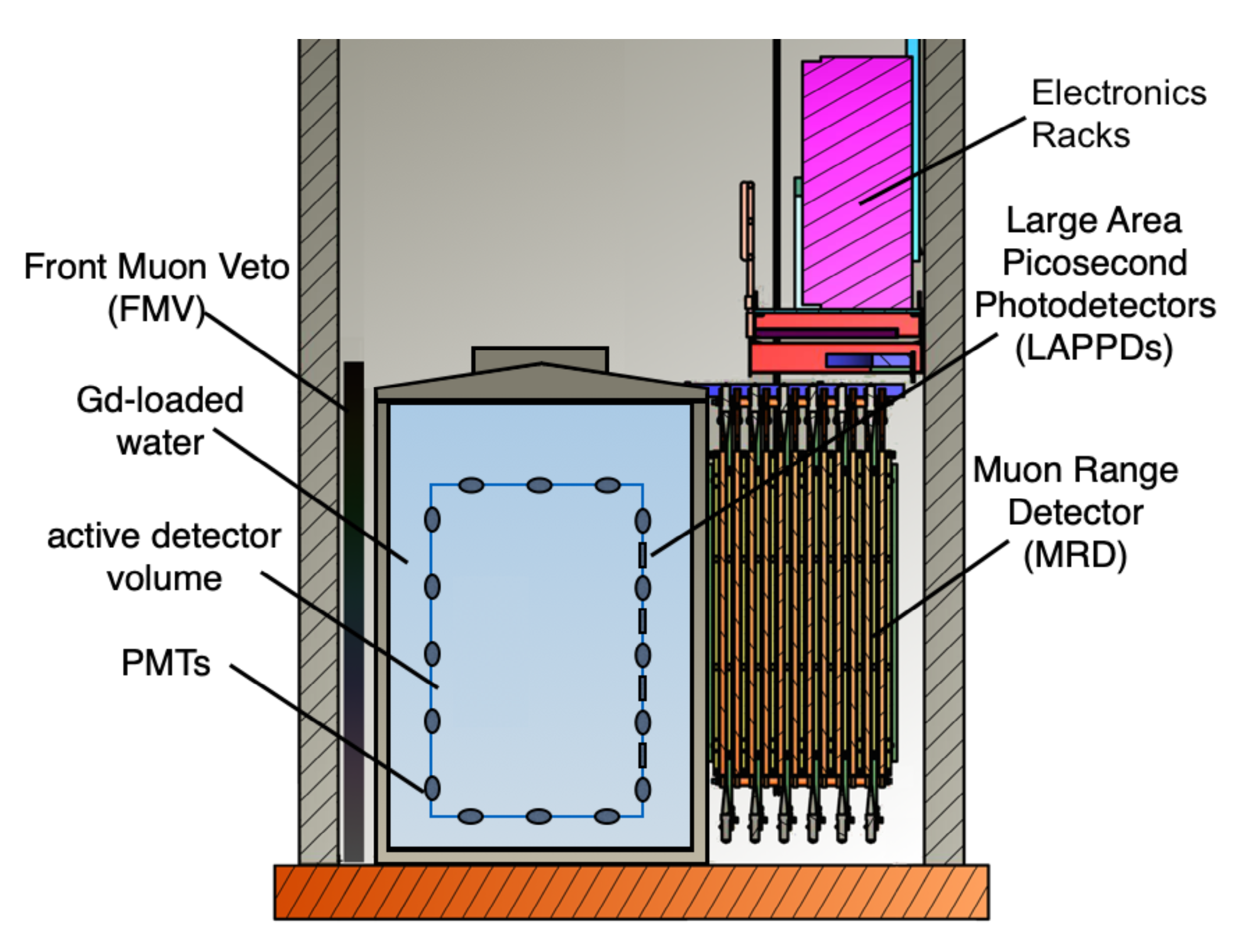}
	\end{center}
	\caption{A schematic drawing of the ANNIE detector system. The neutrino beam enters from the left and then passes through the three major detector subsystems: (i) the Front Muon Veto, (ii) the Gd-loaded target tank instrumented with PMTs and LAPPDs, and (iii) the Muon Range Detector. }
		\label{fig:anniedetector}
\end{figure}

In this section we present a functional summary of the detector components relevant to the reconstruction of neutrino events. The detector consists of three major components, as shown in Figure~\ref{fig:anniedetector}. Note that the neutrino beam enters from the left in this figure. In order of position along the beam these are: (1) the FMV; (2) a gadolinium-loaded water tank containing the neutrino target and optical instrumentation; and (3) the MRD positioned downstream from the neutrino target and consisting of interleaved layers of iron and scintillator.

Each detector element plays a crucial role in event reconstruction. The FMV tags charged products from upstream interactions in the rock, enabling the separation of these external events from those originating in the water tank. Photosensors on the tank walls detect Cherenkov radiation produced by energetic charged particles, providing the majority of information on event kinematics and topology. The downstream MRD stops muons that exit the tank, providing some tracking information and allowing the reconstruction of their energy from their penetration depth. Finally, the addition of gadolinium to the water enhances the visibility of neutron captures, enabling high efficiency neutron counting.

\subsection{Target Tank and Photomultiplier Tube Array}

\begin{figure}[h!]
\centering
\includegraphics[scale=0.45]{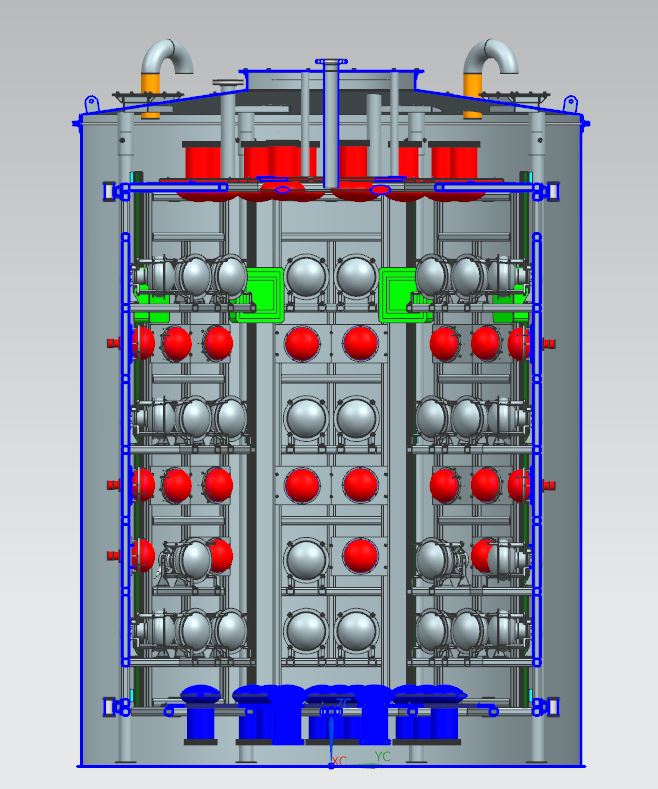}
\hspace{0.2cm}
\includegraphics[scale=0.222]{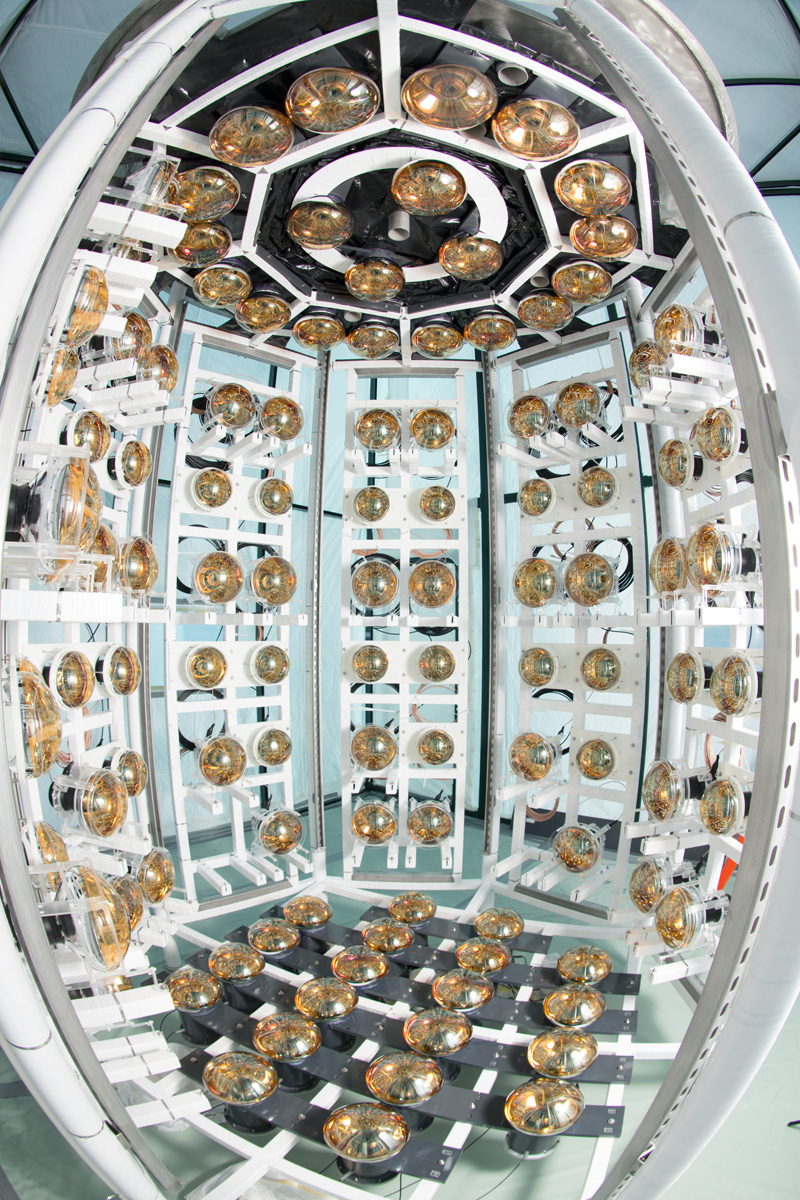}
\caption{(Left) The 3D design of the ANNIE detector and (Right) an image of the inner structure in a clean room environment after all the PMTs have been mounted. Hamamatsu 10-inch PMTs are in blue and grey, Hamamatsu 8-inch PMTs are in red on the sides, and ETEL 11-inch PMTs are in red at the top. The ETEL PMTs were a prototype model that did not go into commercial production. Information on the performance can be found in Reference \cite{etel}. The green rectangles represent notional and not actual LAPPD positions. }
\label{fig:Model_Construction}
\end{figure}

The ANNIE tank is an upright, cylindrical steel storage tank 3.96 meters tall and 3.05 meters in diameter. The inside and outside of the tank are coated with a corrosion-resistant epoxy and a plastic white PVC inner liner is used for compatibility with the gadolinium-loaded water. The tank is sealed with a custom lid that provides access for instrumentation cables, calibration devices, and other test equipment. Sitting inside the tank is a steel inner structure that holds an array of 132 PMTs of three different models, as shown in Table~\ref{anniepmts}. Figure \ref{fig:Model_Construction} (left) depicts the 3D design of the detector, with PMTs on the inner structure grouped by color according to their model.  These provide roughly 14\% photocathode coverage of the inner structure area.

\begin{table}[h!]
\begin{center}
\begin{tabular}{ |l|c|c|l| } 
 \hline
 PMT Type & Diameter (in) & Number & Location\\
 \hline
  Hamamatsu R7081 & 10 & 72 & bottom and sides \\
  ETEL D784UKFLB & 11 & 20 & top \\ 
Hamamatsu R5912-100 & 8 & 40 & sides\\
 \hline
\end{tabular}
\end{center}
\caption{Photomultiplier types used in ANNIE. Multiple types due to the availability of legacy units of the 10-inch and 11-inch models.}
\label{anniepmts}
\end{table}

The tank PMTs allow reconstruction of the vertex and direction of charged particles via the Cherenkov light distinctive ring pattern. Though many large Cherenkov detectors also use the PMT measured timing of the light, ANNIE is too small for this to be effective due to the PMT TTS compared to light travel times. The PMTs are powered by a CAEN SY527 crate with A374P control cards which allow for individually configurable voltage setpoints. To avoid high voltage on the front photocathode submerged in the water a positive operating polarity was selected. Thus, the cables connected to pickoff boxes on the surface which extract the millivolt-scale PMT signal from the high voltage. PMT waveforms are digitized using custom K0T0 4-channel 12-bit 500 MHz ADC cards. The K0T0 cards have a 2 nanosecond time resolution and voltage readout resolution of $2.415V/2^{12}=0.59$ mV per ADC count. 

The tank PMTs have individual pulse shape and timing properties due to there being three different types of tube and the fact that there can be variations in performance even between units of the same type. For the purposes of this study however, tracking is done by pattern recognition only and timing is only used to correlate hits in the target tank, MRD, and FMV for event building and not for reconstruction. Thus, while nanosecond scale timing calibration using a diffuser ball and PILAS picosecond pulsed laser were performed, the details are not relevant here. Similarly, energy estimation is done using muon track length in the water tank and steel layers of the MRD and not by photon counting as is often done in large Cherenkov detectors. Thus knowing the gains of the individual PMTs only serves as an input to the detector simulation and is not directly used to obtain the track energy. Thus both the timing and gain calibration procedures will be discussed in a future detector paper as they are not relevant for this study but will become important after inclusion of LAPPDs into the analysis and when studying neutron capture efficiencies in gadolinium water and WbLS.

\subsection{Muon Range Detector (MRD) and Front Muon Veto (FMV)}

The MRD is an iron-scintillator tracking detector designed to measure the energy and directions of muons originating from neutrino interactions in the tank. The momentum and energy of muons stopped in the MRD can be inferred from the distance they traverse in the steel layers of the subdetector, while their direction can be reconstructed by fitting the muon track recorded by the scintillator paddles. Originally designed for the SciBooNE experiment~\cite{sciboone},  it was modified for ANNIE and now consists of eleven steel plates sandwiched between eleven layers of plastic scintillator paddles, six horizontal and five vertical.  The MRD uses five different PMT types. The horizontal planes consist of 14-stage EMI 9954KB PMTs as well as EMI 9839b and 9939b PMTs. The vertical planes consist of 10-stage Hamamatsu 2154-05 PMTs and RCA 6392A PMTs.

\begin{figure}[h!]
	\begin{center}
		\includegraphics[width=0.46 \linewidth]{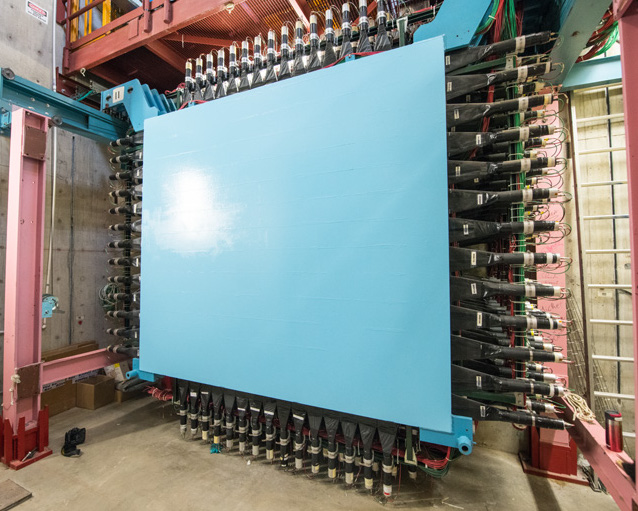} 
     \includegraphics[width=0.45 \linewidth]{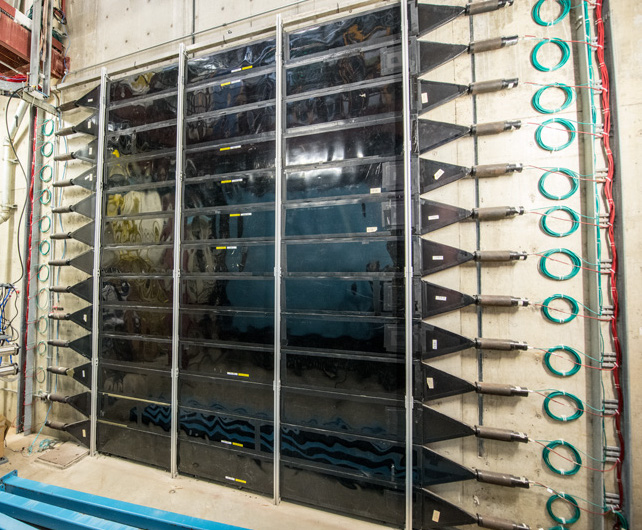} 
  \end{center}
	\caption{(Left) Muon Range Detector: 11 alternating vertical and horizontal layers of scintillating paddles with a 5 cm iron absorber in between the layers to range out muons originating from the neutrino interactions in the tank. (Right) Front Muon Veto: Two layers of scintillator paddles designed to detect and reject muons produced in the dirt upstream of the detector}
		\label{fig:mrdfmv}
\end{figure}

Figure~\ref{fig:mrdfmv} shows a picture of the MRD. The 6 horizontal layers contain 156 of the scintillator paddles while the 5 vertical layers contain another 150 scintillator paddles. The 5-cm iron absorber layers are placed in between the scintillator layers to range out traversing muons. All scintillator paddles are instrumented with 2-inch PMTs read out via LeCroy 3377 Time-to-Digital Converter (TDC) units with 0.5 ns time resolution. 
The efficiencies of the MRD scintillator paddles were determined using samples of muons that traverse the whole MRD both using beam neutrino data and cosmic muon data. The average efficiency for all channels in use was determined to be $(92.1 \pm 7.9)$ \%~\cite{nielsony}. Since muons from neutrino interactions typically have more than four MRD hits, the chance to miss a muon entirely is small, but it does add systematic uncertainty to the measured range.

The FMV is designed to reject muons produced in interactions upstream of the detector. The FMV is mounted on the upstream wall of the hall, as shown in Figure~\ref{fig:mrdfmv} (right). It is composed of 26 scintillating paddles, each with an active area of 31 cm x 322 cm and read out by a 2-inch EMI9815 PMT coupled to one end via an acrylic light guide. The paddles are mounted on the upstream wall of the ANNIE hall in two overlapping layers covering a total area of 4~meters in height by 3.2~meters in width. 

The efficiencies of the FMV paddles were initially measured using cosmic muons in an \textit{ex-situ} laboratory  and determined to be $\bar{\varepsilon}_{\mathrm{FMV}}(\mathrm{lab}) \approx \left(75\pm 7\right)\%$ for all paddles averaged over the active area, but higher on the end with the readout PMT and lower on the opposite end. Since the veto evaluation is performed in an `OR' fashion between the two layers, lower efficiencies at the far end of the paddle in one layer are compensated by higher efficiencies of the corresponding paddle in the neighboring layer. The experiment is hence able to tag entering muons utilizing the combined information from both FMV layers with an efficiency that was found to be $(95.6 \pm 1.6)$\% ~\cite{nielsony} using beam muons detected simultaneously in one layer of the FMV and the MRD.

\subsection{The Booster Neutrino Beam (BNB)}

The Fermilab BNB delivers 8.89\,GeV/c protons in spills with an average rate of 5~Hz to a Beryllium target. The resulting charged particles are then focused with an electromagnetic focusing horn and enter a decay region where they decay into products including neutrinos. Each spill consists of 81 bunches, with each bunch spanning 1.5\,ns and subsequent bunches being separated by 19\,ns. In total, each beam spill lasts 1.6\,$\upmu$s and contains around $4 \times 10^{12}$ Protons On Target (POT). Beam intensity is continually monitored on a spill-by-spill basis by two toroidal current monitors that record proton delivery to within 2\%. 

As configured during ANNIE operation, BNB neutrinos are created predominantly through the decay of positively-charged pions, with small contributions from negative pions, kaons, and the decay-in-flight of secondary muons. The flux exiting the decay region is made up of 93.6\% $\nu_{\mu}$, 5.86\% $\overline{\nu}_{\mu}$ and with a combined $\nu_{e}$/$\bar{\nu}_{e}$ contamination of less than 1\%. Neutrinos are produced with a wide spread of energies, peaking at a little less than 600~MeV. For this paper, the beam simulations from MiniBooNE were adapted for the ANNIE location. A full description of the beam composition from these simulations and a detailed overview of the BNB layout and components can be found in Reference~\cite{AguilarArevalo:2008yp}.

The ANNIE detector is illuminated by a neutrino flux from the BNB of around $2.2\times10^{-8}~\operatorname{cm}^{-2}$ per POT. Over a typical year the total exposure amounts to around $2 \times 10^{20}$ POT, resulting in $\sim$10K/ton Charged Current (CC) neutrino interactions in ANNIE. About 20\% of those CC interactions are expected to produce a muon that will enter and stop in the MRD~\cite{annie}.

\section{Detector Simulation}

\begin{figure}[h]
    \centering
    \includegraphics[width=\textwidth]{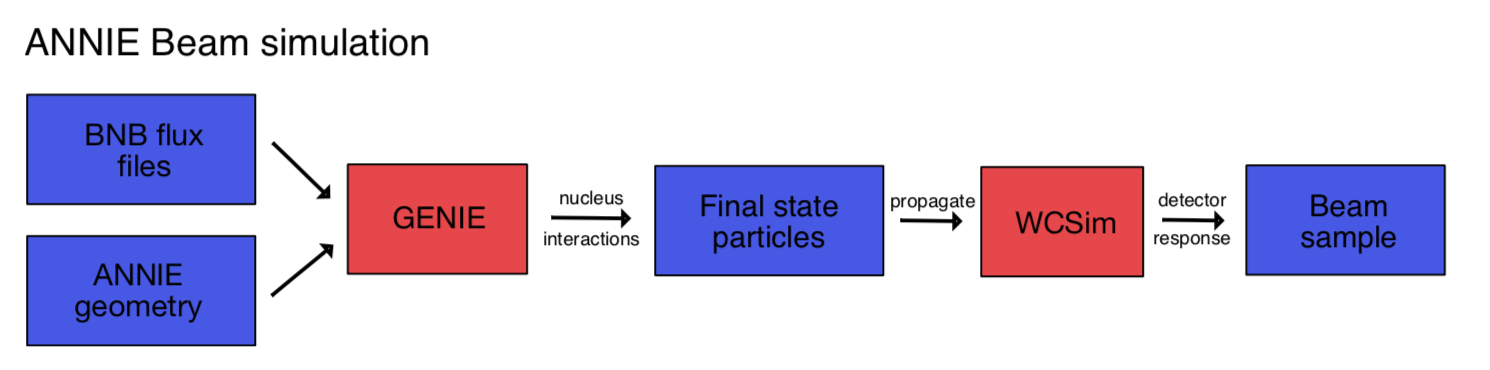}
    \caption{An outline of how beam events are simulated in ANNIE. BNB flux files from SciBooNE and ANNIE's geometry files are read by GENIE. GENIE generates neutrino events, along with information such as positions, daughter particles, energies, and momenta. WCSim then takes this information and propagates the particles throughout the detector and generates the detector response. Output files with the same format as the actual data files are created and can be used for analysis with MC.}
    \label{fig:beam_sim}
\end{figure}

ANNIE uses a series of simulations to develop and test reconstruction techniques to apply to the data, as shown in Figure~\ref{fig:beam_sim}. In the target tank these simulations involve optical processes that determine detector response such as photon propagation and PMT response. Charged particle propagation simulations are used for the MRD, FMV, and in the target tank to take into account energy loss, scattering, and stopping particles such as muons and neutrons. Since the underlying physics for all the above processes are well known, and modeling PMT response is fairly standard and is validated using data, development of tracking algorithms using simulations is a reasonable strategy. The optical simulation of the target tank is validated using the well-known energy spectrum of electrons from muon decays as described below.

It is also desirable to have a simulated sample of neutrino interactions in the target tank in order to have a representative sample of interaction vertices and muon energies and directions. The reconstruction techniques also need to contend with the fact that many CC events will have more than simply a single-track muon. Emitted pions, protons, and neutrons will also be present and make the muon track and energy reconstruction more complex.

\subsection{Neutrino Event Generation}

ANNIE uses the neutrino flux predictions for the Booster Neutrino Beam calculated by the MiniBooNE Collaboration based on simulations of proton-beryllium interactions, followed by propagation of the resulting particles using a Geant4 simulation~\cite{AguilarArevalo:2008yp}. This simulated flux data is then input into a \texttt{ROOT} and \texttt{C++} based neutrino event generator called \texttt{GENIE} (Generates Events for Neutrino Interaction Experiments)~\cite{genie}.
\texttt{GENIE} predicts the properties of all particles that are produced in the primary neutrino interaction, such as the vertex, direction, and energy. 
For the particle properties, \texttt{GENIE} requires the parameters of the neutrino beam flux distribution and information about the detector geometry and material composition.

\subsection{Simulation of Physics Processes}

Once the neutrino interaction and resulting particles are generated, they are propagated through the detector with WCSim~\footnote{ WCSim and associated documentation are publicly available at http://github.com/WCSim}, a \texttt{Geant4}~\cite{geant4a, geant4b, geant4c} based water Cherenkov simulation framework used to develop and simulate large water Cherenkov detectors. 

WCSim was written to simulate water Cherenkov detectors such as Hyper-Kamiokande and thus it was relatively simple to adapt to ANNIE which has a smaller size, different geometry, and different PMTs than Hyper-Kamiokande. A detailed simulation of the ANNIE detector electronics was also added. 
Also added to the simulation were LAPPDs, the external FMV and MRD detectors, and accommodation of the three different PMT types used in the target tank with asymmetrical placement.
More details about the complete detector implementation can be found in Reference~\cite{nielsony}.

ANNIE WCSim uses \texttt{Geant4} version 10.2 configured with the \texttt{FTFP\_BERT\_HP} physics list recommended for high energy physics applications to model physics processes.
More specifically, the \texttt{High Precision (HP)} part of the list is necessary to more accurately model hadronic transport and capture processes for neutrons in low kinetic energy region (<20~MeV). 
Neutron interaction properties are taken from the G4ENDL4.5 data set, which is derived from the data in the ENDF/B-VII.1 library \cite{endf}.
The resulting predictions have been validated against both experiment and the MCNPX simulation package~\cite{marcus}.

\subsubsection{WCSim Optical Model}
WCSim can track the interactions of each particle through the detector, including the production of Cherenkov and scintillation photons, their absorption, reflection, and scattering. 
The optical characteristics of material surfaces in the water tank are modeled with the UNIFIED framework in WCSim as a starting point before tuning.
Photons striking a PMT may be reflected, be absorbed, or generate a photoelectron. The PMT response is parametrized based on empirical data. Our implementation of WCSim does not explicitly simulate the physical processes occuring inside PMTs.

PMT efficiencies are applied in two stages: an initial cut during photon generation uses the maximum quantum efficiency of any sensor in the detector, reducing the number of photons that need to be tracked. 
On each hit a second cut is applied if necessary to account for the difference in quantum efficiency of the particular sensor being hit. 
The charge for each photon is pulled from a random distribution based on the calibration measurements. 
Digitization cuts are then made on simulated data to model multiplicity triggering and charge collection.

\subsubsection{WCSim Tuning}

To improve our understanding of physics processes, we tune the models used in simulation to match what is seen in data.
In ANNIE, a comparison of charged-current interactions in Monte Carlo (MC) and in beam data shows that the total charge distribution in simulation needs to be scaled up to match the charge distribution in data. 
Two parameters that contribute to the total charge seen in the detector are the charge collection efficiency (CE) of the PMTs and the reflectivity of the materials inside the water tank. 
Both needed to be increased from the {\it a priori} nominal values from Super-Kamiokande since the total charge seen in MC was overall less than that seen in data. 

To tune the WCSim framework so that it better represents data, a sample of Michel electrons produced from neutrino-generated stopped muons that entered the tank from the outside was used to evaluate the overall light collection in the detector, while a sample of through-going muons was used for reflectivity studies. Note that to understand the procedures described below it is necessary to understand the concept of PMT {\it clusters}. Particles generating Cherenkov light in the target tank are identified by looking for clusters of PMTs coincident in time. A cluster is defined to be at least 5 PMT hits registered in a time window of 50 ns. 

Since the Michel electron energy spectrum is well known, it was used to adjust the simulation parameters so that simulated charge distributions match the data. The requirements for a Michel electron candidate are: a stopped muon in the tank (FMV hit, but no MRD activity), an electron candidate in the few microseconds following the muon decay, a charge balance (CB) cut that measures the isotropy of the emitted light to filter out through-going muons and other backgrounds near the wall~\cite{pershing}, and a cut to select events with the vertex in the downstream region of the target tank. This last cut is made to further reduce through-going muons since they have a majority of the PMT deposited charge in the downstream region of the water tank.

The PMT charge collection tuning factor, $\text{r}_{\text{CE}}$, is simply a multiplicative factor applied to the nominal charge collection efficiency implemented in WCSim such that $\text{CE}_{\text{tuned}}(\lambda) = r_{\text{CE}} \cdot \text{CE}_{\text{nominal}}(\lambda)$.

\begin{figure}[h]
\centering
\includegraphics[width=0.45\textwidth]{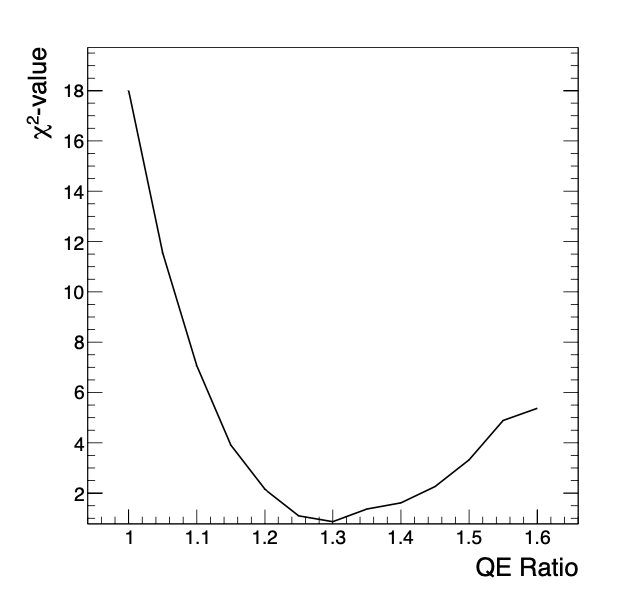}
\includegraphics[width=0.48\textwidth]{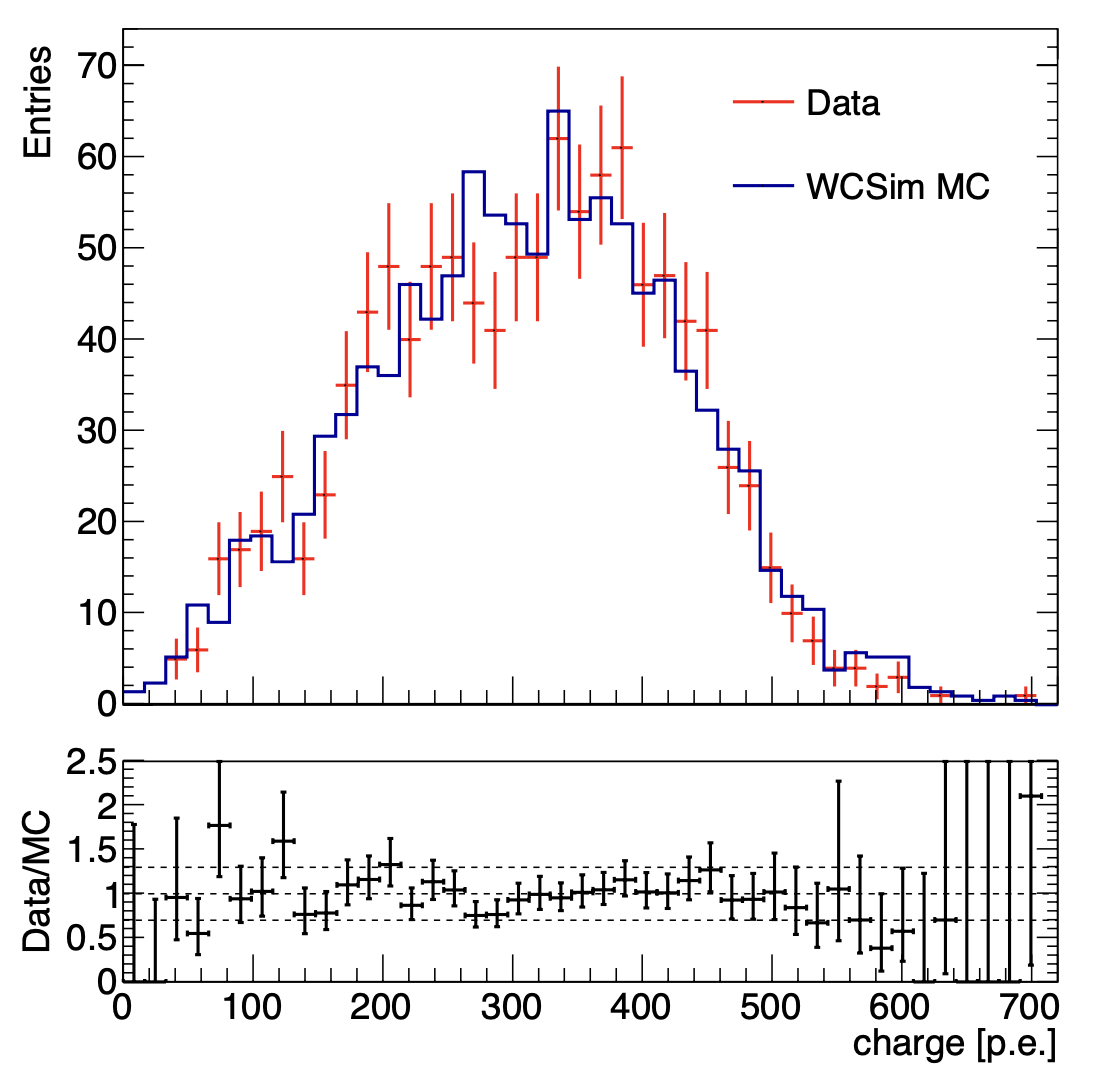}
\caption{(Left) Scan of $\chi^2/DOF(=40)$ value versus $r_{CE}$. (Right) comparison of Michel electron deposited p.e. for Data and WCSim after tuning with $r_{CE}=1.3$.} 
\label{michel}
\end{figure}

The parameter $\text{r}_{\text{CE}}$ was varied in increments of $\Delta r_{\text{CE}}$ = 0.05, in the range [1.0,~1.6], until the $\chi^2/DOF$-value was minimal in comparing the data and simulated charge distribution histograms.
(see Figure~\ref{michel} left). Figure~\ref{michel} shows final agreement of data with the simulation after tuning with the value $r_{CE}=1.3\pm 0.1$. 

There are several physical  reasons for $\text{r}_\text{{CE}}$ to vary from the nominal value of 1.0. These include: (i) the simulation of the PMT shape in WCSim is simply a truncated sphere instead of the actual shape of individual PMT types, which effects the photon reflection geometry, (ii) WCSim does not allow for photons entering the glass bulb to create p.e.s, and (iii) the waveform integration procedure was not identical in the beam data processing and single p.e. determination. This last factor will be corrected after reprocessing all the ANNIE data.

Tuning of the reflectivity is done in a similar manner as the tuning of PMT charge collection efficiency. The data set used are nearly horizontal through-going muons generated from neutrino interactions in the rock and dirt between the BNB decay pipe and the FMV. The forward-going direct Cherenkov light from such muons deposits light mainly on the downstream PMTs, while upstream PMTs see mainly reflected light. Thus, the ratio between the upstream and downstream light is very sensitive to the reflectivity of the detector components. Note that ANNIE is too small for absorption in the water to have any significant effect on the optics.

To select through-going muons for this analysis we require a coincidence between the FMV, target tank, and MRD.   The reconstructed muon track in the MRD was then extrapolated to a plane in front of the FMV. In the simulation muons are placed at that particular extrapolated position and are initialized in the direction of the reconstructed MRD track. While the real through-going muon sample is comprised of a variety of energies, the behavior in the tank will be quite similar for all muons as essentially all of the BNB-generated muons are minimum ionizing particles. Thus, the energy of the simulation muons was set to 2 GeV to ensure that the muons reliably pass through the entire tank and also produce a track in the MRD. The topology of the reconstructed MRD track in the simulation can then be validated against the original MRD track that was observed in the data to ensure that the same event properties are replicated between data and simulation.

Several components of the water tank contribute reflections but three dominate in terms of surface area: the teflon-wrapped inner steel structure, the black plastic sheet surrounding the inner structure, and the PMT photocathodes. Other surfaces, such as the PMT holders are much smaller in area and thus nominal wavelength-independent values (0.9 for the 10-inch Hamamatsu side holders and 0.7 for the bottom 10-inch Hamamatsu bottom and ETEL top holders) for their materials are assumed. Similarly, the black plastic sheet used in ANNIE is similar to that used in Super-Kamiokande and so the shape of the measured values as a function of wavelength are adopted for ANNIE tuned to be in the range of 0.11 to 0.14.

\begin{figure}
\centering
\includegraphics[width=0.8\textwidth]{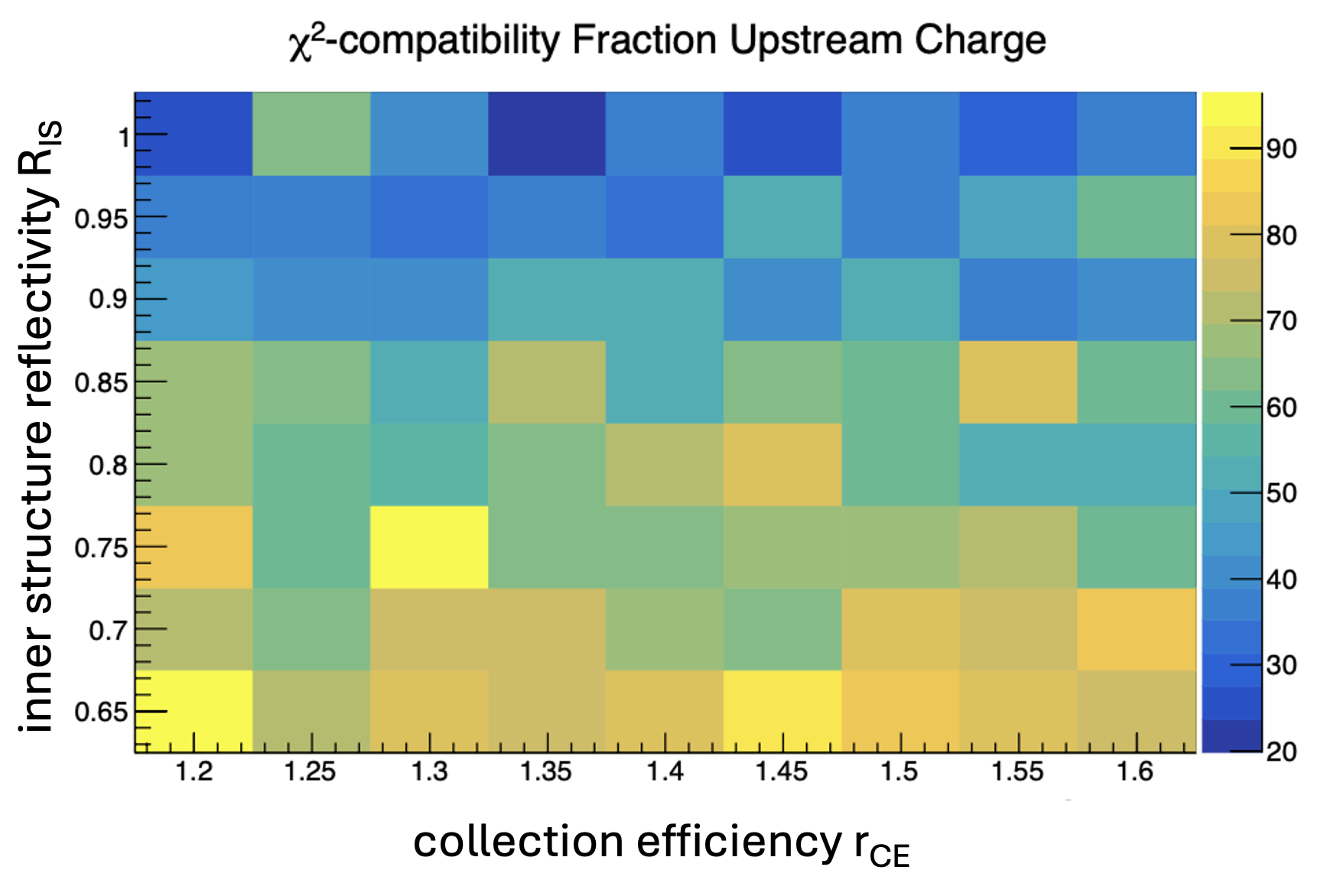}
\caption{Comparison between data and simulation of the through-going muon upstream/downstream light ratio  efficiency with various values of the collection efficiency and inner structure reflectivity.}
\label{ris_scan}
\end{figure}

After the black sheet the inner steel structure  has the largest surface area. Thus, its reflectivity ($\text{R}_\text{{IS}}$) was tuned first in a manner similar to that of the PMT efficiency.
$\text{R}_{\text{IS}}$ was varied in steps of $\Delta \text{R}_{\text{IS}}$ = 0.05, while also varying $\text{r}_{\text{CE}}$ by the same step size to determine the optimal combination of tuning factors. 
The fraction of charge seen in the upstream part of the tank is compared to the fraction of charge seen downstream.
A reflectivity of $R_{\text{IS}}$ = 1.00 was optimal fairly independent of  $\text{r}_\text{{CE}}$ as shown in Figure~\ref{ris_scan}. This value represents 100\% reflectivity for teflon, which is somewhat larger typical measured values of >90\% in our optical region of interest. This high value likely due to the fact that we did not adjust the reflectivity of the black plastic sheet from the Super-Kamiokande nominal values, which may be an underestimate. Both teflon and black plastic sheet reflectivity parameters are nearly degenerate in their effects so we chose to vary only one of them.

For a more detailed discussion on the tuning procedure see Reference~\cite{nielsony}.

\section{MRD Tracking}
\begin{figure}
    \centering
    \includegraphics[width=1.0\textwidth]{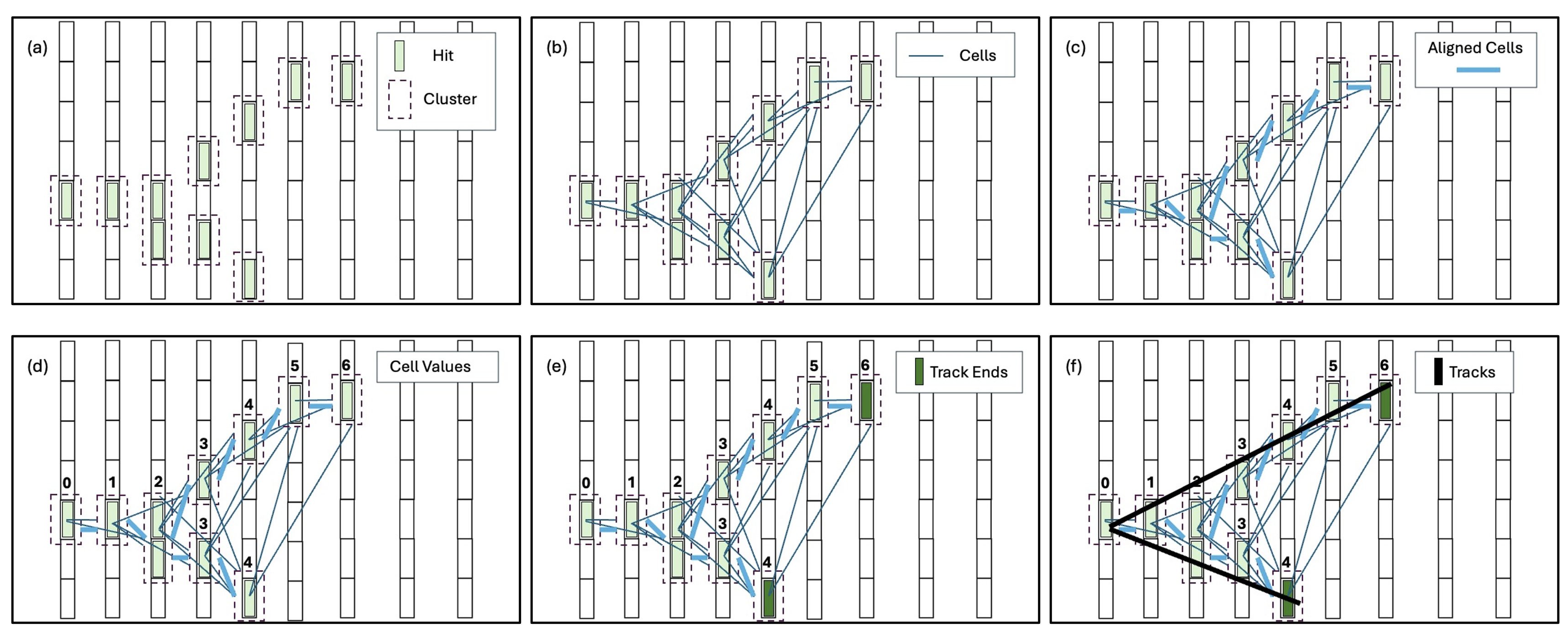}
    \caption{A diagram depicting the stages of track reconstruction: (a) Hits are grouped into clusters; (b) Cells are formed from the clusters, representing track segments; (c)-(d) Cells are paired with upstream and downstream neighbors, with statuses assigned to each pair to determine order; (e) Track ends are identified and tracks are reconstructed; (f) Linear fit to identified tracks.}
    \label{fig:mrd_track_reco}
\end{figure}

The MRD is used to fit the direction and location of muon tracks exiting the target tank. This information is then combined with Cherenkov image reconstruction in the tank to form a complete reconstruction of the neutrino-induced muon vertex and track from CC neutrino interactions. MRD track fitting comes from the positional information inherent in the crossed orthogonal scintillator paddle pattern as shown in Figure~\ref{fig:mrdfmv}.

The MRD track reconstruction algorithm is based on the cellular automation method used by K2K \cite{k2k_track} and T2K \cite{t2k_track, kikawa}. Horizontal and vertical views of the track are reconstructed independently and then combined in the end. The steps of this process are shown in Figure~\ref{fig:mrd_track_reco}. The steps are:
(a) For each layer, hits in adjacent paddles within a 30 ns window are grouped into a cluster; (b) for each cluster, make cells by combining it with all clusters in the next two downstream layers (for example the third cluster from the left has four associated cells); (c) For each cell, assign an upstream neighboring cell based on a linear least squares 2-D fit on the paddle centers with a squared residual of <125 cm2. If multiple upstream cells are available, the one with the smallest residual is selected. Thus in figure 7(c) the far right cluster has two cells as shown by the blue lines; (d) Each cell is then assigned a value depending on the number of upstream aligned cells. Thus the far right cluster has a value of 6 as there are six aligned upstream cells; (e) Cells that are not the upstream neighbor of a cell with a higher value define the ends of tracks; (f) Identified 2-D horizontal and vertical tracks are now passed on to an algorithm that will combine these into 3-D tracks.

The horizontal and vertical tracks are then matched based on a figure of merit calculated using the proximity of track start and end points, agreement between the two views on which half of the MRD was penetrated, and the steepness of the track from the beam (shallower tracks are preferred). 
Unmatched 2-D tracks are discarded. 
For the resulting 3-D tracks, a linear fit is performed on all the clusters to determine the final track position and direction.

\begin{figure}[h!]
\centering
\includegraphics[width=0.7\textwidth]{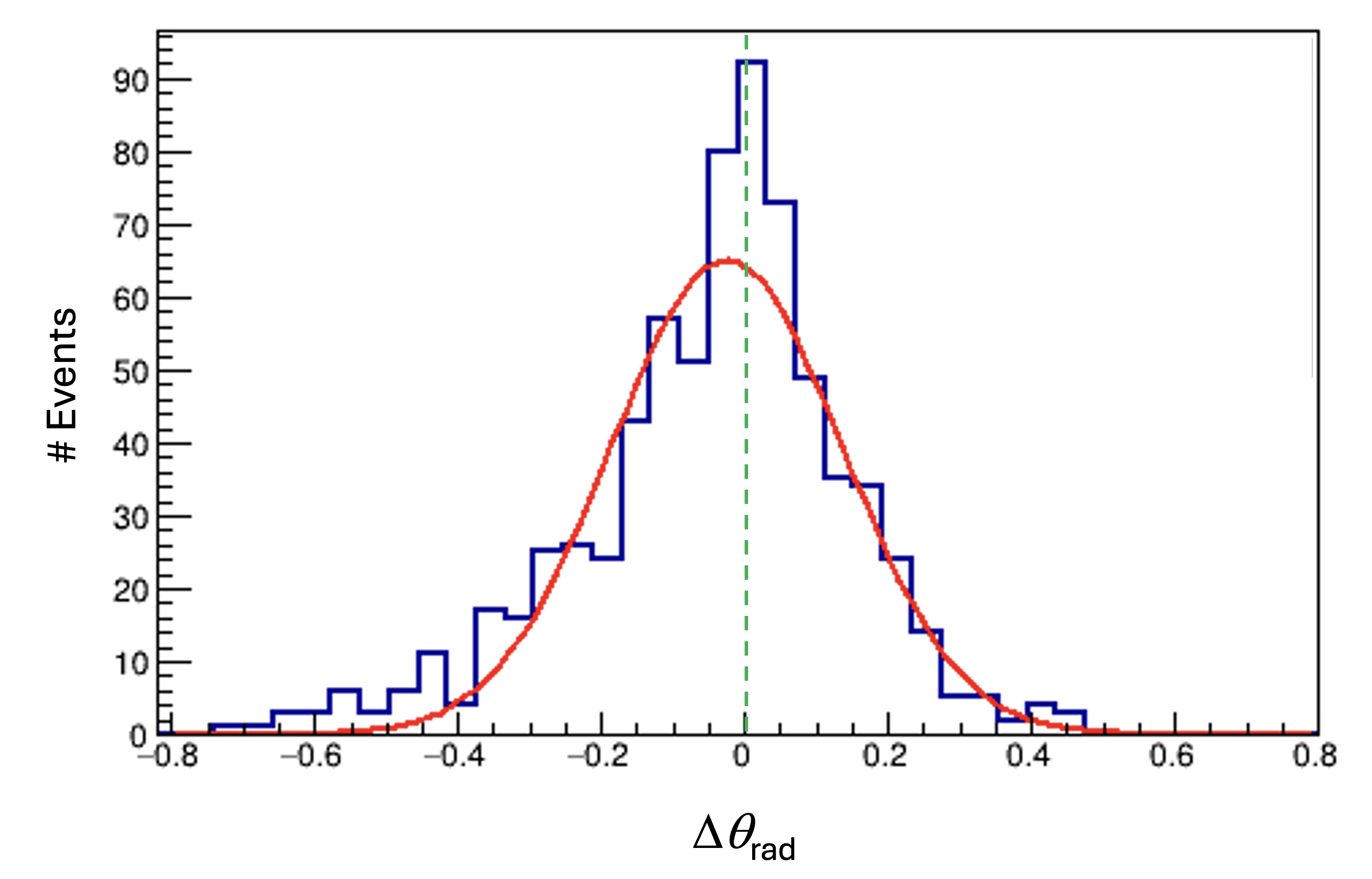}
\caption{Angular difference ($\Delta \theta$) in one plane between the fitted MRD direction and the truth information for a sample of simulated beam events after making all cuts. The sigma of the Gaussian fit (shown in red) is 9.3 degrees.}
\label{fig:mrdres}
\end{figure}

The tracking algorithm was tested on a sample of muons simulated from neutrino events that originate inside a right cylinder of radius and height of 1 meter located at the center of the target tank. This volume defines a 3.1 ton Fiducial Volume (FV) inside the target tank. As shown in Figure~\ref{fig:mrdres}, the detector simulation gives an angular resolution of 9.3$^\circ$ in one plane for this algorithm, which corresponds to 13.2$^\circ$ in three dimensions.  More details of the MRD track-fitting algorithm can be found in References~\cite{marcus, nielsony, julie}. 

\section{Event Reconstruction} 

A full reconstruction of neutrino interactions requires that the interaction vertex in the target tank be determined. In order to ensure that interactions in the target tank can be accurately reconstructed, it is necessary to restrict the usable sample to vertices that occur within the FV.  In addition, the muon track length in the tank must be taken into account when determining the total energy of the muon track via energy loss per unit length. The analysis presented here uses the MRD muon track extrapolated back into the target tank. It then moves along the track using the pattern of PMT detected light to image the edge of the Cherenkov ring, thus determining the event vertex. Finally, the total muon track path length is used to determine the muon track energy with corrections for multiple scattering and small variations in energy loss due to changes in muon energy along the track. 

\subsection{Event Selection}

The BNB neutrino data presented here was collected from February 21, 2024 to April 13, 2024. For this analysis it was desirable to have a sample of CC events with a single identified muon track with little or no other activity. To do this, only events occurring within 2 $\mu$s of the start of the 1.6 $\mu$s long BNB spill were analyzed. The following cuts were then applied:\\
\begin{enumerate}
\item Required no hits in the FMV to reject events with vertices upstream of the experimental hall.
\item Required one and only one track identified in the MRD.
\item Required that the MRD and tank event times are coincident with each other within 50 ns.
\item Required that the back-extrapolated MRD track pass within 1 meter of the center of the target tank volume as an initial FV cut.
\item Required that the muon track stop in the MRD in order to allow energy determination.
\end{enumerate}

These cuts yielded a sample of 995 neutrino event candidates. Identical cuts applied to the simulated data resulted in a sample purity of 91\% charged current $0\pi$ events. 

\subsection{Track Reconstruction}

The MRD reconstruction constrains the neutrino interaction vertex in the target tank to lie close to the extrapolated track, but does not determine the exact vertex location. To do this it is necessary to use the tank PMT array to image the distinctive Cherenkov ring edge at candidate locations on the back-extrapolated MRD track. This ring imaging method has two steps: (1) for every hit PMT calculate the track point that Cherenkov light would have been emitted in order to generate that hit, and (2) assuming that Cherenkov light emitted from a Minimum Ionizing Particle (MIP) is constant, find the point on the track where the corrected detected light has a large change, indicating the location of the Cherenkov ring edge and consequently the event vertex.

\begin{figure}[h!]
    \centering
    \includegraphics[width=0.75\textwidth]{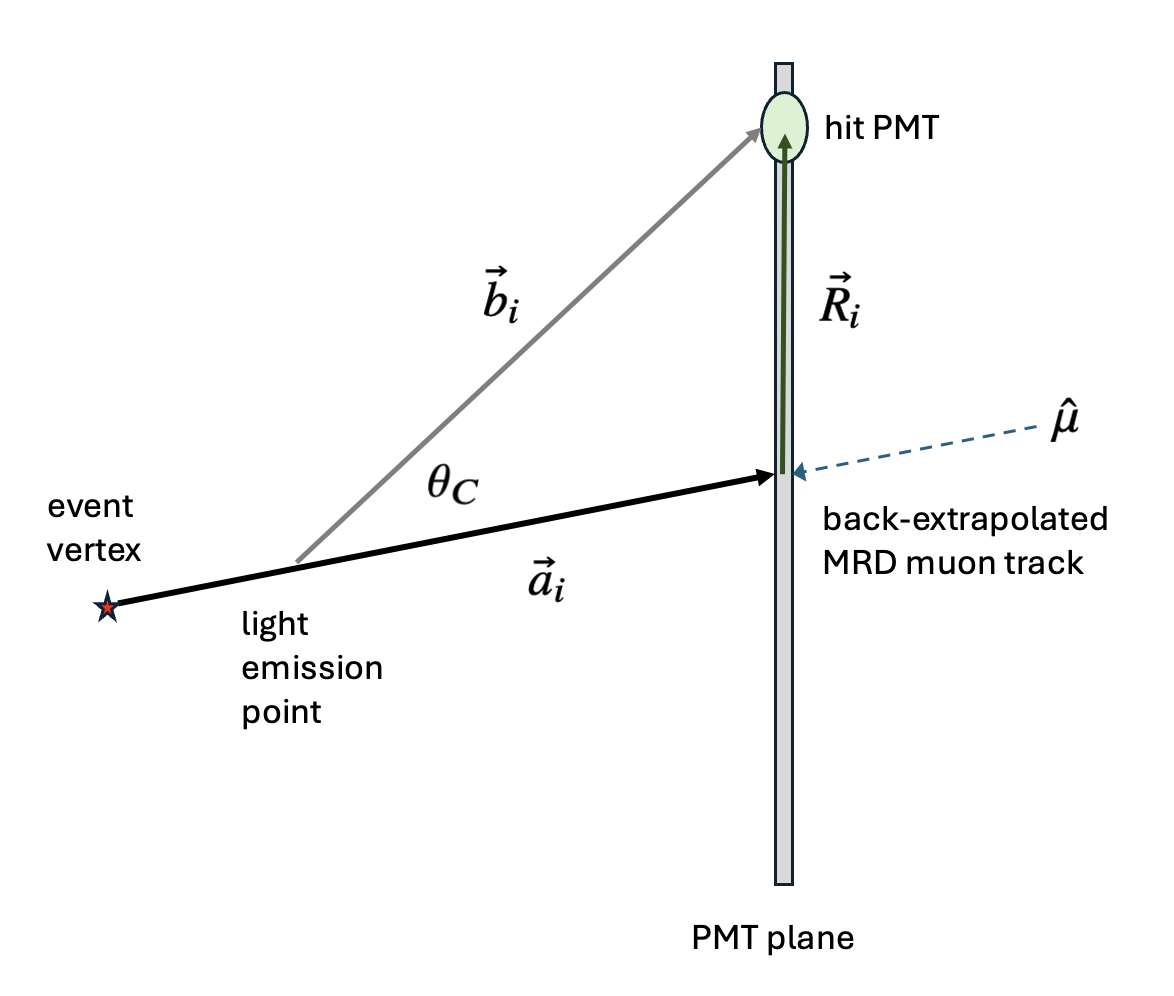}
    \caption{Cherenkov geometry used to determine the light emission point ($\vec{a}_i$) of the $i^{th}$ hit PMT. The unit vector $\hat{\mu}$ is defined as the direction opposite to the muon direction.}
    \label{ring_img1}
\end{figure}

The first step is accomplished by using the well-known Cherenkov geometry as shown in Figure~\ref{ring_img1}. For every $i^{th}$ hit PMT there is a mapping to the light emission  point on the MRD track direction constrained by the Cherenkov angle ($\theta_C=42^o$) in water. We define $\vec{a}_i$ to be the vector from this emission point to the point where the muon track exits the PMT plane. $\vec{R}_i$ is the vector from that exit point to the center of the hit PMT, and $\vec{b}_i$ is the vector from that emission point to the hit PMT.  We desire to know the vector $\vec{b}_i$ in terms of known quantities $\vec{R}_i$ and $\hat{\mu}$ (the MRD track direction) so that the number of measured photoelectrons can be corrected for both the spreading of the Cherenkov ring with distance and the PMT geometrical acceptance. This can be done by referring to the figure and performing simple vector algebra, noting that $\vec{a}_i=-a_i\;\hat{\mu}$:\\

\begin{equation}\label{eq:arb}
    \vec{a}_i + \vec{R}_i = \vec{b}_i \implies  \vec{a}_i = \vec{b}_i - \vec{R}_i
\end{equation}

\begin{displaymath}
\hat{\mu}\cdot\vec{a}_i=\hat{\mu}\cdot\vec{b}_i-\hat{\mu}\cdot\vec{R}_i
\end{displaymath}

\begin{equation}\label{eq:ai}
    a_i = b_i~\cos{\theta_C} + \left( \hat{\mu} \cdot \vec{R}_i\right )
\end{equation}
\\
Taking the cross product of $\hat{\mu}$ with  Equation~\ref{eq:arb} and solving for $b_i$:\\

\begin{equation}\label{eq:bi}
    \hat{\mu} \times \vec{a}_i = 0 = \left(\hat{\mu} \times \vec{b}_i\right) - \left(\hat{\mu} \times \vec{R}_i\right)\implies  b_i = -\frac{| \hat{\mu} \times \vec{R}_i |}{\sin{\theta_C}}
 \end{equation}
\\ 
Substituting Equation~\ref{eq:bi} into Equation~\ref{eq:ai} yields:\\
\begin{equation}
    a_i = -\frac{| \hat{\mu} \times \vec{R}_i|}{\tan{\theta_C}} + \left(\hat{\mu} \cdot \vec{R}_i\right)
\end{equation}
\\
Finally, substituting Equations~\ref{eq:ai} and \ref{eq:bi} into Equation~\ref{eq:arb} gives $\vec{b}_i$ in terms of $\vec{R}_i$ and $\hat{\mu}$:\\
\begin{equation}
\vec{b}_i=\left[ \frac{| \hat{\mu} \times \vec{R}_i|}{\tan{\theta_C}} - \left(\hat{\mu} \cdot \vec{R}_i\right) \right] \hat{\mu}+\vec{R}_i
\end{equation}

\begin{figure}[h!]
    \centering
    \includegraphics[width=0.6\textwidth]{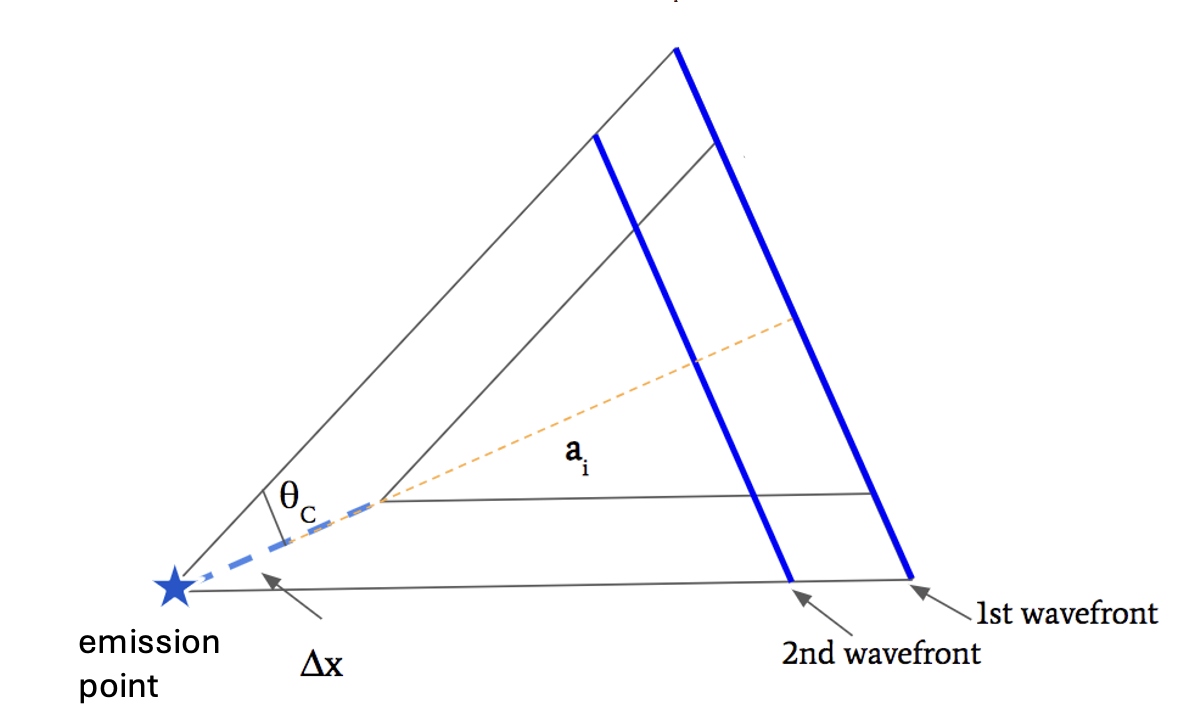}
    \caption{Expansion of the Cherenkov wavefront from a track segment $\Delta x$ with distance decreases the photon density.}
    \label{fig:eta_defn}
\end{figure}

Knowing $\vec{b}_i$ enables the geometrical correction procedure as follows. For a MIP the number of Cherenkov photons emitted per unit distance is roughly constant, which we define as ($\eta$) with units of photons/cm. Referring to Figure~\ref{fig:eta_defn}, as the photons from track segment $\Delta x$ expand with distance from the emission point, the area ($A_{ring}$) of the illuminated Cherenkov ring increases and thus the photon density decreases.
This $ A_{ring}$ potentially overlaps with some PMTs at an acceptance angle whose cosine is given by $-\hat{b}_i\cdot\hat{n}_i$, where $\hat{n}$ is the normal to the PMT axis.
We define $f_i$ as the fraction of $A_{ring}$ that the $i^{th}$ PMT subtends, corrected for the PMT acceptance angle. Summing up $f_i$ over all hit PMTs in $A_{ring}$ gives the total fraction of the area occupied by photocathode surface. Then, dividing the measured total number of p.e.'s that fall in $A_{ring}$ by $f$ should ideally yield a constant $\eta$. This simple picture ignores late-arriving reflected light (which can be partially cut by timing)  and also Cherenkov light from any charged hadron activity at the neutrino interaction vertex, but it is nevertheless a good approximation to what is seen in the data and simulation.

\begin{figure}[h!]
    \centering
    \includegraphics[width=1.00\textwidth]{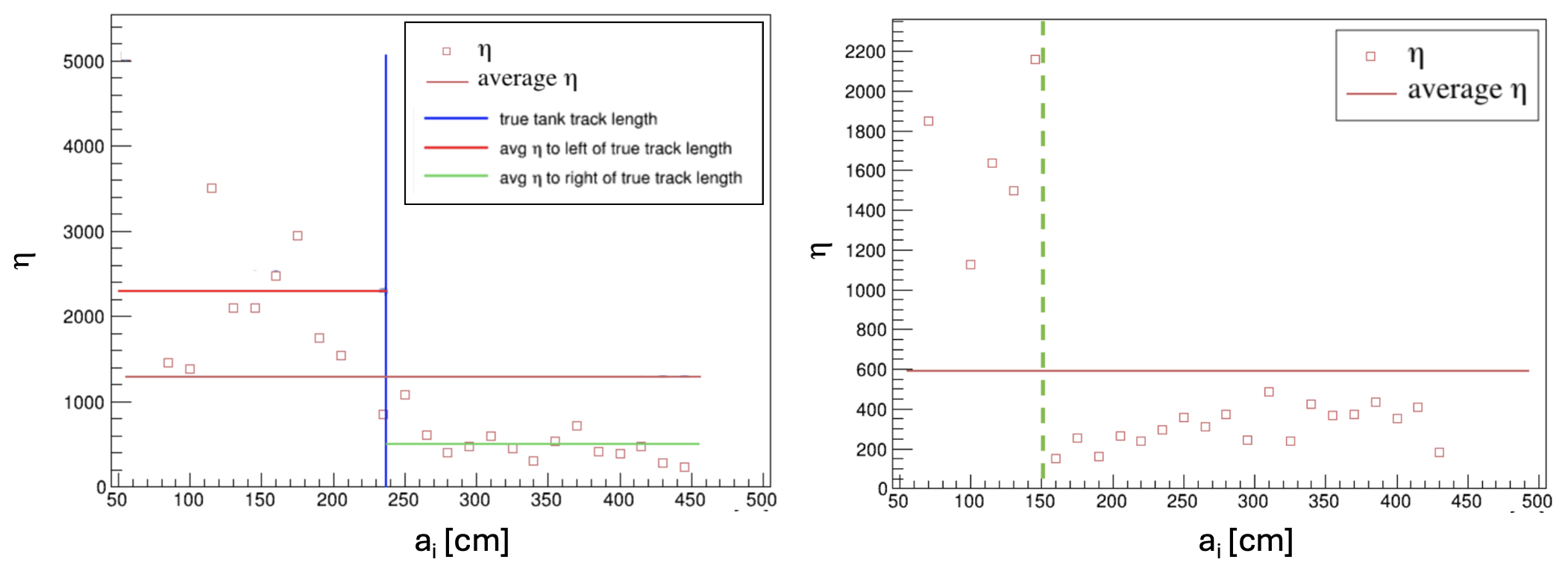}
    \caption{A measurement of $\eta$ versus $a_i$ for a simulated beam event (left) and a data event (right).}
    \label{eta12}
\end{figure}

\begin{figure}[h!]
    \centering
    \includegraphics[width=0.99\textwidth]{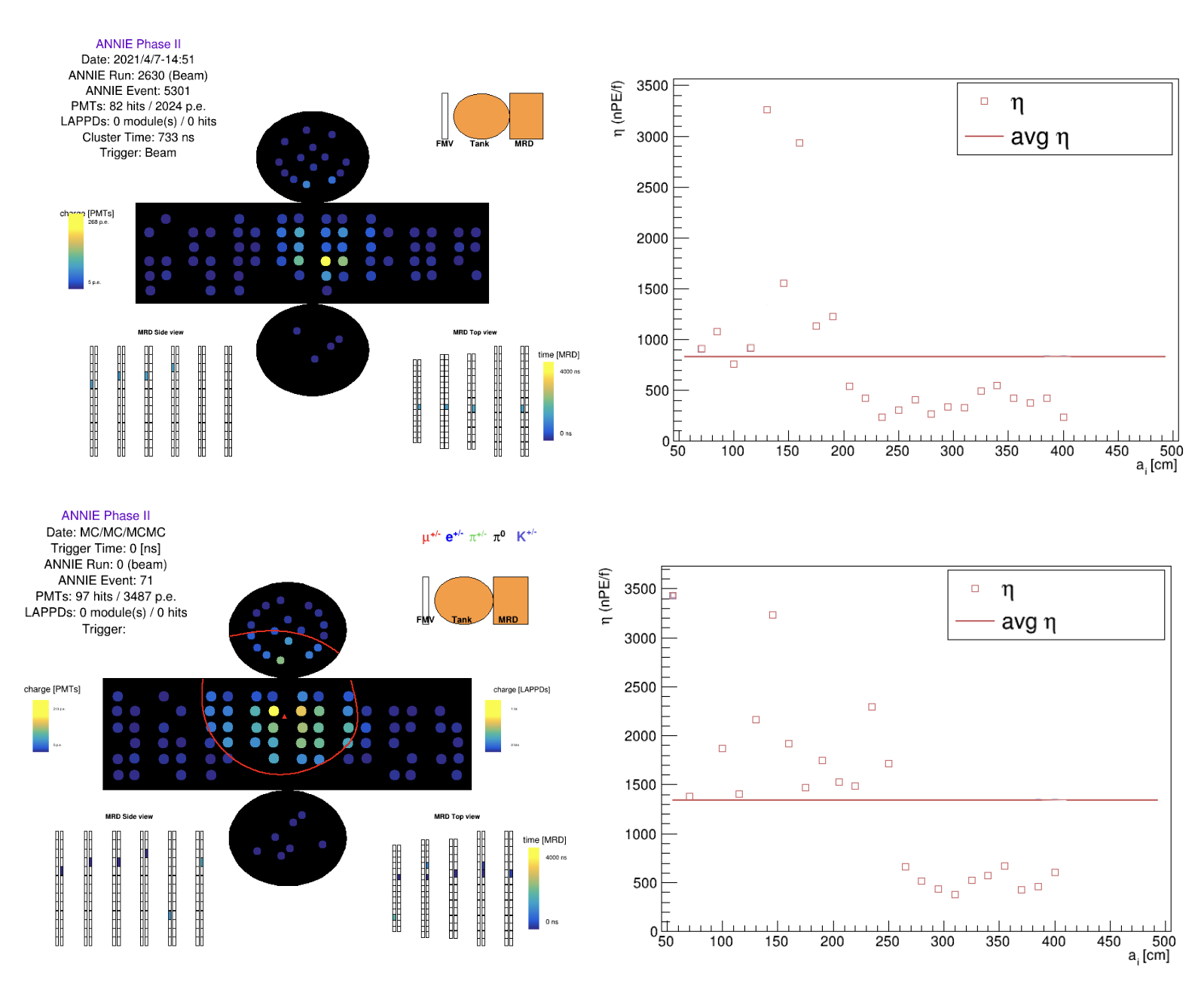}
    \caption{(Top) An event display showing a a beam event alongside the $\eta$ versus $a_i$ plot. (Bottom) a simulated event with similar geometry. The red ring on the event display represents the truth information location of the Cherenkov ring.}
    \label{eta3}
\end{figure}

Figure~\ref{eta12} (left) shows a simulated neutrino event that produced a single muon track that met the event selection cuts.  Each small rectangle shows the value of $\eta$ in 10 cm steps of $a_i$ in the direction along the MRD track in the tank ($i$ in this case stands for the index of the track distance, not the hit PMT). The value of 10 cm was selected as a value that would have significant hit PMT statistics but yet not be so large as to affect the precision of the final fit.  The blue vertical line indicates the MC truth location of the ring edge. To the left of the line at smaller $a_i$ values the PMTs are inside the Cherenkov cone, whereas to the right they are outside the cone. The green and red horizontal lines are the average values of $\eta$ to the right and left of the blue line, respectively.  The right figure shows a similar plot for a data event. In this case the vertical green dashed line shows an obvious break that signifies the likely ring edge. Outside the ring, the reflected light level is approximately constant as expected, whereas inside the ring there is a larger value of $\eta$ but there is also increased variability due to the increasing smaller number of PMTs hit as the track approaches the wall. Thus, the strategy is to devise an algorithm to optimize where to draw the ring edge on the $\eta$ versus $a_i$ plot.

To further illustrate this point, Figure~\ref{eta3} shows a data event (top) and a simulated event (bottom) with similar geometry. Next to each $\eta$ versus $a_i$ plot is an unrolled event display that shows the light level in individual hit PMTs. Note that outside the cone $\eta$ is fairly flat, but inside the cone the values are much larger with more variability. These features were used to fit the location of the vertex using a basic Recurrent Neural Network (RNN) trained on simulated data. The RNN uses a linear layer to convert the hidden parameters to a single value output. There is only one hidden layer and only four neurons. The loss function is MSE (Mean Squared Error) and no normalization method is used. This is a relatively standard algorithm that is in wide use.

\begin{figure}[h!]
\centering
\includegraphics[width=0.65\textwidth]{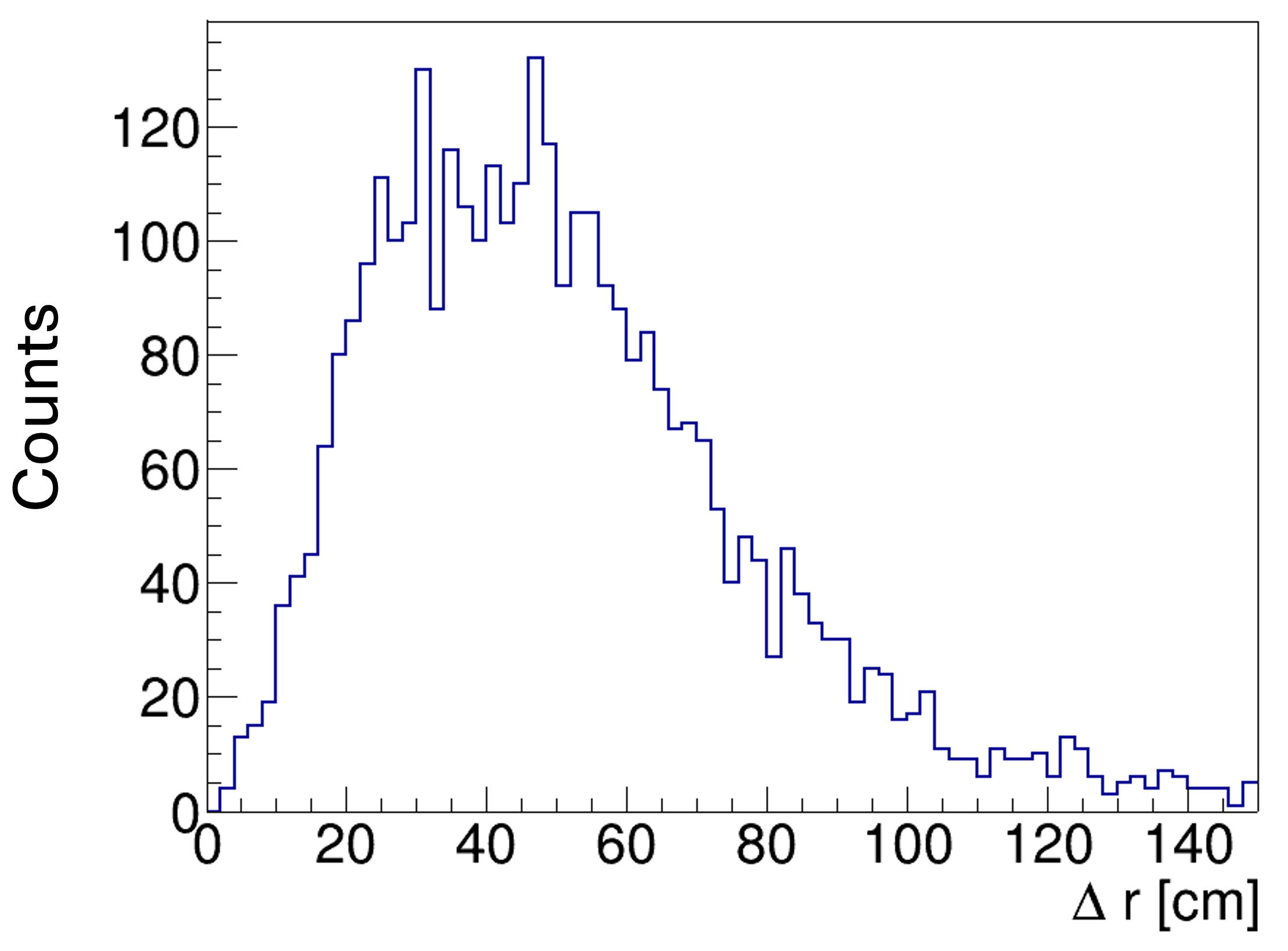}
\caption{Distribution of the absolute value of the difference between the true and reconstructed muon vertex for a sample of simulated beam events in 2 cm bins. The 68\% inclusion value is 60 cm.}
\label{fig:vertex}
\end{figure}

To train the RNN model using simulated data the $\eta$ values calculated at evenly spaced (every 10~cm) track length intervals for each candidate event were saved to an output file along with the true track length of the muon. Half of the dataset was used to train the model while a quarter of the data was used for testing. About 1,200 candidate events were used as inputs to the model. The RNN is given 10,000 epochs to minimize a cost value and to reach a cost value where the training and validation datasets converge. The final model was saved as an output and used to determine the track lengths of each candidate muon in actual data. The track lengths in turn were used to determine the vertex. Figure~\ref{fig:vertex} shows the distribution of the difference between the true and reconstructed muon vertex for a sample of events that pass all cuts and reconstruct within the FV. The 68\% inclusion distance is at 60 cm, which is worse than that achieved in large detectors such as Super-Kamiokande~\cite{skrecon} (about 16 cm) due to the small tank and large size of the PMTs (20 -- 28 cm) compared to the distances involved (100-200 cm). For these large detectors however, all the ANNIE events would be short exiting tracks with vertices at the edge of their fiducial volumes so the comparison is not straightforward. This is the motivation for ANNIE to deploy and utilize LAPPDs to be able to more precisely track muons.  These devices have both fast timing and the ability to record the photon hit location  to better than 1 cm.

\subsection{Muon Energy Determination}
The muon energy is calculated from the fitted track lengths in the water tank and MRD combined with the well-known values of the stopping power ($-dE/dx$) as a function of energy in water and iron. At the sub-GeV energy of BNB muons nearly all energy loss is via ionization characterized by the Bethe-Bloch equation~\cite{pdgbb}.

For a muon with $\mathcal{O}(10^2~\text{MeV})$ kinetic energy the stopping power is roughly 2 MeV/cm in water and 11 MeV/cm in iron. For example, a track length in the tank of 140~cm and an average $dE/dx$ of 2~MeV/cm results in 280~MeV of energy deposited in the water, while an effective track length of 40~cm and an average $dE/dx$ of 11~MeV (assuming iron is the main source of energy loss) results in 440~MeV deposited in the MRD.
In total, this muon is estimated to have 720~MeV of kinetic energy. This is very simple approximation. 
To make a more accurate determination an iterative procedure that takes into account both the energy dependence of the stopping power and muon multiple scattering is used, as outlined below: 
\begin{enumerate}
    \item Use the nominal stopping power described above  to make an initial estimate $E_0$ of the energy deposited in the tank and in the MRD.
    \item Use this $E_0$ to calculate a new $dE/dx$ value from the Bethe-Bloch equation.
    \item Starting from the vertex, multiply the new $dE/dx$ value by the fixed amount of distance traveled by the muon (2 cm in our case). Add the deposited energy to a cumulative sum.
    \item Subtract the deposited energy from $E_0$ and input the updated energy $E^{'}_{0}$ into the Bethe-Bloch formula to get a new $dE/dx$ value.
    \item Repeat the above two steps until the remaining input energy falls below 20~MeV. 
    \item Add the remaining input energy ($\sim$20~MeV) to the cumulative sum.
\end{enumerate}

\begin{figure}[h!]
\centering
\includegraphics[width=0.95\textwidth]{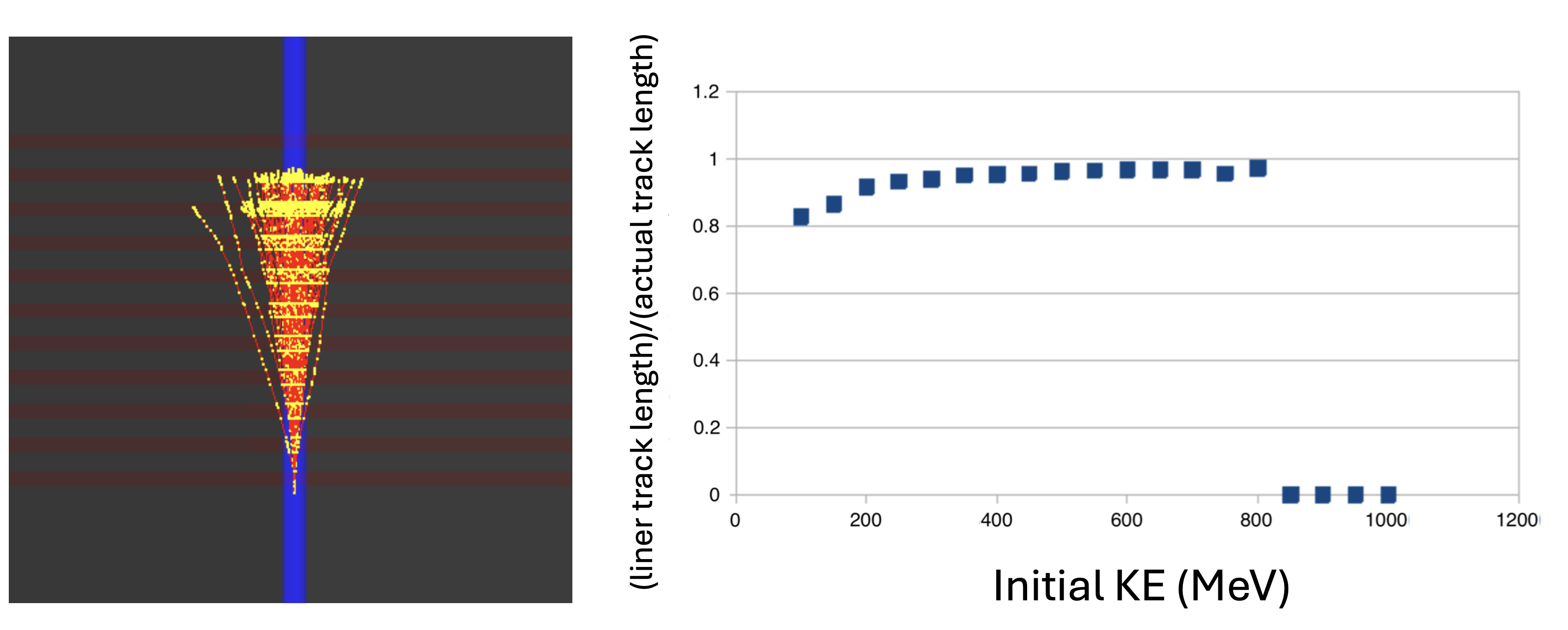}
\caption{(Left) Tracks followed by simulated 500 MeV kinetic energy muons in an MRD-like geometry using Geant4. The increase in track length due to scattering must be taken into account on average when determining the energy loss. (Right) The ratio between the straight line track fit and the average track length of simulated muons. This factor is used to correct the energy based on the initial estimate of muon energy. Above 800 MeV no muons stop, so there is no correction applied.}
\label{scattering}
\end{figure}

This procedure assumes the muon track is a straight line. This is not exactly correct, as the high density of iron increases the chances of scattering of the muon as it traverses the MRD as illustrated in Figure~\ref{scattering} (left). This scattering adds to the total track length and results in more energy deposited as compared to a straight linear track. To account for this a simulation was used to calculate energy-dependent correction to the total track length (and hence the energy). This method  counts the number of MRD iron plates crossed by the fitted track. This count is converted into an effective amount of iron traversed by the muon by multiplying by 5 cm (thickness of iron plate) and dividing by the cosine of the fitted track angle. Since it is uncertain where the muon stops in the final iron layer, 2.5 cm is added to the final effective track length.

The average energy-dependent impact of scattering is then calculated using a GEANT4 simulation of muons traversing iron. For muons of all energies in the range of 100 -- 1,000 MeV, the actual track length traversed is longer than the track length estimated by calculating the distance between the MRD entry and stopping or exit points in the current ANNIE reconstruction methods. Impacts of scattering are taken into account by using a correction factor, as shown Figure~\ref{scattering} (right).

\begin{figure}
    \centering
    \includegraphics[width=0.90\textwidth]{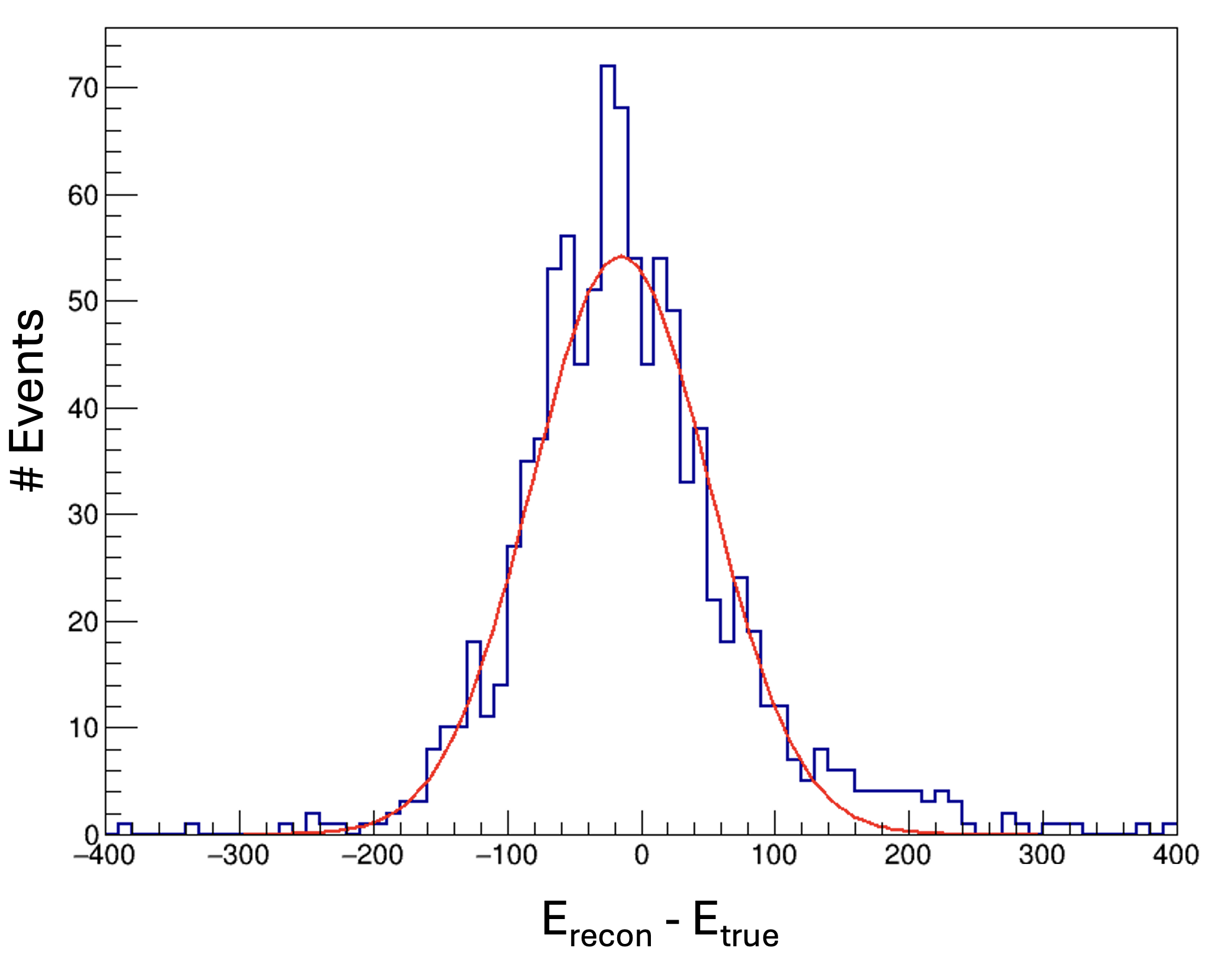}
    \caption{A comparison of reconstructed muon energy to the true energy of the muon for a sample of simulated beam events. A Gaussian curve is fitted to the sample of events after a fiducial cut is made on the reconstructed muon vertices. The mean energy difference is about -15~MeV, meaning the energy reconstruction algorithm is slightly underestimating the muon's kinetic energy. The sigma of the fit (shown in red) is about 66 MeV. Events with a large positive difference are mainly attributed to events with unresolved pions.}
    \label{fig:h_true_reco_ediff}
\end{figure}

\begin{figure}
    \centering
    \includegraphics[width=0.90\textwidth]{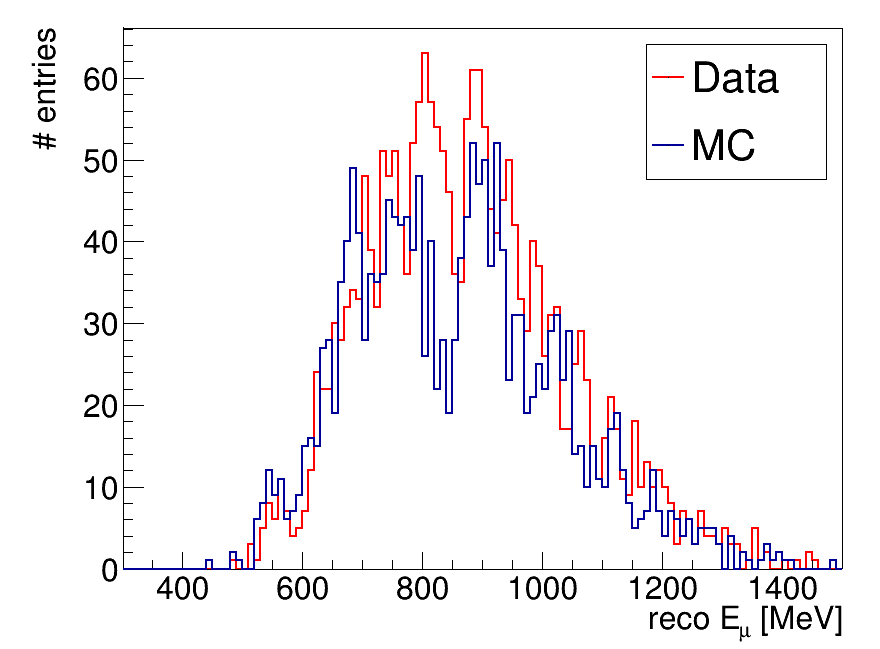}
    \caption{Comparison of the reconstructed energy of simulated BNB events with a similar number of ANNIE data events.}
    \label{fig:compare_data_mc_energy}
\end{figure}

Figure~\ref{fig:h_true_reco_ediff} shows a distribution of the difference between the reconstructed muon energy and the true muon energy for simulated beam events whose vertices are within the 3.1 ton FV. 
The sigma of about 66 MeV corresponds to an energy resolution $\Delta E / E$ of about 10\% (assuming the average muon energy peak is $\sim$800~MeV). Finally, Figure~\ref{fig:compare_data_mc_energy} shows a comparison of the muon energy distribution of simulated sample of beam events versus a similar number of ANNIE data events, verifying that the reconstruction technique presented in this paper reproduces the expected BNB spectrum for CC0$\pi$ events. 

\section{Conclusions}

The basic ANNIE detector consists of an upstream muon veto, a 3-meter diameter tank containing Gd-doped water and a 2.3 meter diameter PMT array, and a downstream tracking muon spectrometer. ANNIE is unusual in that it is small compared compared to the light travel time smearing due to PMT timing resolution, thus the convention water Cherenkov vertex reconstruction using mostly timing does not work well. In addition, ANNIE event vertices all occur within 2 meters of a downstream tank wall, which in large water Cherenkov detectors are typically rejected as outside the reconstruction fiducial volume.  In this paper we demonstrated that the combination of the MRD and PMT array pattern recognition could be combined together to enable fitting single muon neutrino interactions that would simply be rejected as outside the fiducial volume in larger detectors and not used for physics analysis
This will allow ANNIE to make preliminary cross section measurements of charged-current events that generate a muon that stops in the MRD. 

ANNIE has now installed LAPPDs on the downstream wall of the target tank, which will have time resolutions on the order of 60-70 ps. A follow-on paper to this one will describe reconstruction using these novel detectors for the first time, and will compare LAPPD-enhanced reconstruction over the more conventional methods presented here.

\acknowledgments

This work was supported by: 
(1) the U.S. Department of Energy (DOE) Office of Science under award numbers: DE-SC0018974 (UC Berkeley), DE-SC0009999 (UC Davis), DE-SC0015684 (Iowa State University), DE-SC0024684 (Florida State University), and DE-SC0014223 (South Dakota Mines);  (2) the National Nuclear Security Administration, Office of Defense Nuclear Nonproliferation  Research and Development (DNN R\&D) under contracts: DE-AC02-05CH11231 (LBNL), DE-AC52-07NA27344 (LLNL), DE-AC02-98CH10886 (BNL),  and Nuclear Science and Security Consortium award number DE-NA0003180 (UC Davis); (3) Deutsche Forschungsgemeinschaft grants 456139317 (Hamburg), 490717455 (Mainz and Tübingen), and 552099472 and RTG2796 “Particle Detectors” (Mainz);
(4) the U.S. National Science Foundation grant numbers: PHY-2310018 (Ohio State) and PHY-2047665 (Rutgers); (5) the U.K. Research and Innovation (UKRI) [MR/Y034082/1]
(Warwick); and (6) T\"{U}BA, T\"{U}B\.{I}TAK and Scientific Research projects (BAP) of Erciyes University in T\"{u}kiye under the grant numbers
FBAU-2023-12325, FOA-2025-15098, and FBAU-2025-14357. The LLNL release number is LLNL-JRNL-2013004.

Finally, we gratefully acknowledge all the Fermilab scientists and staff who supported this
work through their technical expertise and operational assistance at the Booster Neutrino Beam.

% Bibliography

%% [A] Recommended: using JHEP.bst file

\bibliographystyle{JHEP.bst} 
\bibliography{biblio}

\end{document}